# M3Scope: a 3D multimode multiplane microscope for imaging nanoscale dynamics in soft matter

Steven Huysecom[1,5], Francisco Bevilacqua[3], Roger Bresolí-Obach[2], Sudipta Seth[1], Luis M. Liz-Marzán[3,4,7,8], Dagmar R. D'hooge[5], Johan Hofkens[1,6], Susana Rocha[1,*], Boris Louis[1,*]

[1]KU Leuven, Department of Chemistry, Molecular Imaging and Photonics, B-3001 Leuven, Belgium

[2]Department of Analytical and Applied Chemistry, AppLightChem, Institut Quimic de Sarria, Universitat Ramon Lull, Via Augusta 390, 08007 Barcelona, Catalunya, Spain

[3]CIC biomaGUNE, Basque Research and Technology Alliance (BRTA), 20014 Donostia-San Sebastián, Spain

[4]Centro de Investigación Biomédica en Red, Bioingeniería. Biomateriales y Nanomedicina (CIBER-BBN), 20014 Donostia-San Sebastián, Spain

[5]Ghent University, Laboratory for Chemical Technology, B-9052, Zwijnaarde, Belgium

[6]Max Planck Institute for Polymer Research, 55128 Mainz, Germany

[7]Ikerbasque, Basque Foundation for Science, 48009 Bilbao, Spain

[8]CINBIO, University of Vigo, 36310 Vigo, Spain

Corresponding Author
* boris.louis@kuleuven.be , susana.rocha@kuleuven.be

# Abstract

Fast, volumetric imaging that integrates multiple imaging modalities is essential for probing dynamic, heterogeneous soft and biological matter. Here, we present the M³Scope, a simple yet versatile multiplane microscope that extends widefield detection with a modular multimode cube to enable dual-color fluorescence, polarization-resolved, and correlative brightfield–fluorescence imaging while (i) preserving simultaneous 3D acquisition at high frame rates (~100 fps) and (ii) requiring minimal realignment. We demonstrate its potential by investigating polymer dynamics across multiple spatial and temporal scales. In an acrylamide type polymerizing network, dual-color tracking of 100 nm and 300 nm fluorescent probes revealed size-dependent viscosities diverging from ~15 mPa*s to ~40 mPa*s after gelation. Polarization-resolved rotational tracking of gold bipyramids yielded viscosities within ~10% of theoretical values in 80–95% glycerol and remained accurate in the high-viscosity regime (≥400 mPa*s) where translational motion is undetectable. Fluorescence–brightfield imaging correlated structural changes during poly(isopropyl acrylamide) (pNIPAM) phase separation as well as variations in tracer diffusivity, linking morphology and dynamics in 3D. Taken together, these results show that the M³Scope delivers high-speed volumetric imaging with flexible modality switching, providing a powerful platform for studying dynamic, heterogeneous systems across disciplines, from polymer physics to cell mechanobiology.

# Introduction

Modern-day microscopy has revolutionized imaging in many areas of life sciences, material sciences, physics, biology, and chemistry, by enabling the visualization of molecular structures and dynamic processes inside soft and complex matter [1–9]. Biological systems and soft matter, such as cells, tissues, and polymers, exhibit intricate structures and highly dynamic behavior across multiple spatial and temporal scales. As these structures and processes are inherently three-dimensional (3D-), the development and application of advanced 3D microscopy methods have become essential for a comprehensive understanding of the structural organization and dynamic heterogeneity [10,11].

Several optical strategies have been developed to achieve volumetric imaging. In confocal microscopy, a 3D image is reconstructed by scanning a focused laser beam across every pixel in the sample volume. This point-by-point acquisition provides axial sectioning but suffers from low temporal resolution due to its sequential nature. In contrast, widefield microscopy captures an entire two-dimensional plane instantaneously, allowing for high acquisition speeds, although with limited axial resolution. In widefield systems, 3D positions of single emitters can be retrieved through point spread function (PSF) engineering, where the optical response of the microscope is modified to encode the emitter's axial position. This strategy, often employed in single-molecule localization microscopy, enables 3D super-resolution imaging by determining emitter positions with nanometric precision, well below the diffraction limit. This approach enables 3D localization over axial ranges of up to ~4 µm [12–14]), but is inherently limited to low emitter densities, as overlapping PSFs cause localization ambiguities. To bridge the gap between the axial sectioning of confocal microscopy and the speed of widefield imaging, light-sheet microscopy has been developed. By illuminating the sample with a thin laser sheet instead of a single focal point, it improves temporal resolution, though 3D reconstruction still requires scanning, which limits the acquisition speed. [15]

Multiplane imaging overcomes this constraint by enabling instantaneous volumetric acquisition without PSF engineering. By splitting the emitted light, introducing controlled differences in the optical path length and projecting the resulting focal planes onto distinct regions of the detector, multiplane systems achieve high-speed volumetric imaging, reaching frame rates of up to ~100 fps, over volumes of several hundreds of cubic micrometers, making them well-suited for particle tracking at relatively high tracer densities and speeds. [16–21]

While advanced volumetric techniques have expanded the capabilities of optical microscopy, measurements performed in a single imaging mode often provide only limited insight into complex soft and biological materials. The observed dynamics in these systems typically arise from the interplay of multiple molecular, chemical, and structural factors. For example, cellular organization and motion can be influenced simultaneously by cytoskeletal remodeling, membrane tension, and local biochemical signaling, whereas in polymeric or colloidal systems, particle interactions, viscoelasticity, and phase heterogeneity collectively determine the dynamic behavior [22–24](**2020 Tang**). Capturing this complexity requires high-content, multiplexed imaging, in which multiple targets or processes of interest are visualized simultaneously and distinguished within the same sample. In practice, multiplexing is often obtained at the cost of complicated experimental setups and/or sequential acquisition of different channels, reducing time resolution. Despite that some multiplane systems have been extended to multi-color or phase imaging these implementations typically require substantial optical modifications and limited flexibility in switching between imaging modes. [18,25,26] As a result, the widespread application of multiplexing in fast 3D imaging remains restricted, despite its importance for the comprehensive characterization of complex systems [21].

To address this, we introduce M³Scope, a simple yet powerful optical design that extends multiplane systems to enable multimode imaging, while maintaining 3D acquisition and high temporal resolution. By incorporating only a few additional optical elements, the system allows rapid, switching between imaging modes, providing a highly versatile platform for high-speed volumetric imaging. In its current implementation, the system readily supports dual-color fluorescence, 3D polarization imaging, and simultaneous brightfield–fluorescence imaging and can easily be extended to other modes. To illustrate these capabilities, we applied it to polymer physics, where fast, volumetric, and multimode imaging is particularly valuable. Polymers display complex, time- and scale-dependent dynamics that can be probed by tracking tracer particles embedded in the network. From their Brownian motion, local mechanical properties such as viscosity and viscoelastic moduli can be inferred, a technique known as passive microrheology [28–30]. Because polymer systems are often heterogeneous and non-ergodic, high-throughput, 3D tracking of many particles is required to capture their full mechanical response. [31,32]

Here we show that the M³Scope enables direct access to these spatial and temporal heterogeneities by combining fast volumetric imaging with flexible multimode detection. First, Dual-color imaging is used to track tracer particles of different sizes (100 and 300 nm), in a PAA polymerizing network, revealing size-dependent viscosities ranging from ~15 mPa*s to ~40 mPa*s after gelation. This heterogeneity would be occulted in single-channel measurements. Second, we demonstrate polarization-resolved imaging, capturing full 3D rotational dynamics of anisotropic probes in high viscosity regimes where translational tracking is not applicable. Finally, we also show how the M³Scope can integrate multiple contrast mechanisms within a single setup. Combining brightfield and fluorescence imaging, structural transitions during poly(N-isopropyl acrylamide) (pNIPAM) phase separation, a well-established reference system in the field of thermoresponsive polymers [33], were correlated with changes in tracer diffusivity, linking morphology and dynamics in a single experiment. The versatility of M³Scope makes it broadly applicable to a wide range of experimental contexts, from soft-matter systems to live-cell imaging, bridging physical and biological studies of dynamic, heterogeneous systems.

# Results

*Fast 3D multimode imaging platform with 16 outputs*

The M³Scope is based on a previously described custom-built multiplane wide-field design [10] (**Fig. 1A**), consisting of an epifluorescence microscope equipped with five laser lines (405, 532, 561, 640, and 1064 nm), a brightfield lamp, two CMOS cameras, and a multiplane prism that splits the emitted light into detection paths of different optical lengths, projecting a series of focal planes onto the two detectors (**Fig. 1B** and detailed in patent EP3049859A1).

This way, simultaneous images from eight distinct axial planes are obtained, of which four planes are projected onto each camera. The detection arm is constructed as an 8f system, creating an additional infinity space where optical elements can be introduced. To enable multimode imaging, a multimode magnetic cube (MMC) is placed in this infinity space. Inserting a dichroic or beamsplitter into the MMC creates a second detection arm, which mirrors the primary path. Mirrors M2 and M3 direct this secondary beam such that it enters the multiplane prism from a different face with a slight vertical offset relative to the primary path (**Fig. 1A**). The tube lenses TL1 and TL2 create the image onto the cameras, with their distances from the prism precisely matched to ensure that both channels image an equivalent set of focal planes. The second detection arm results in sixteen simultaneously acquired planes, where each plane in the second channel is positioned below its counterpart in the first channel at the same focal depth (**Fig. 1B**).

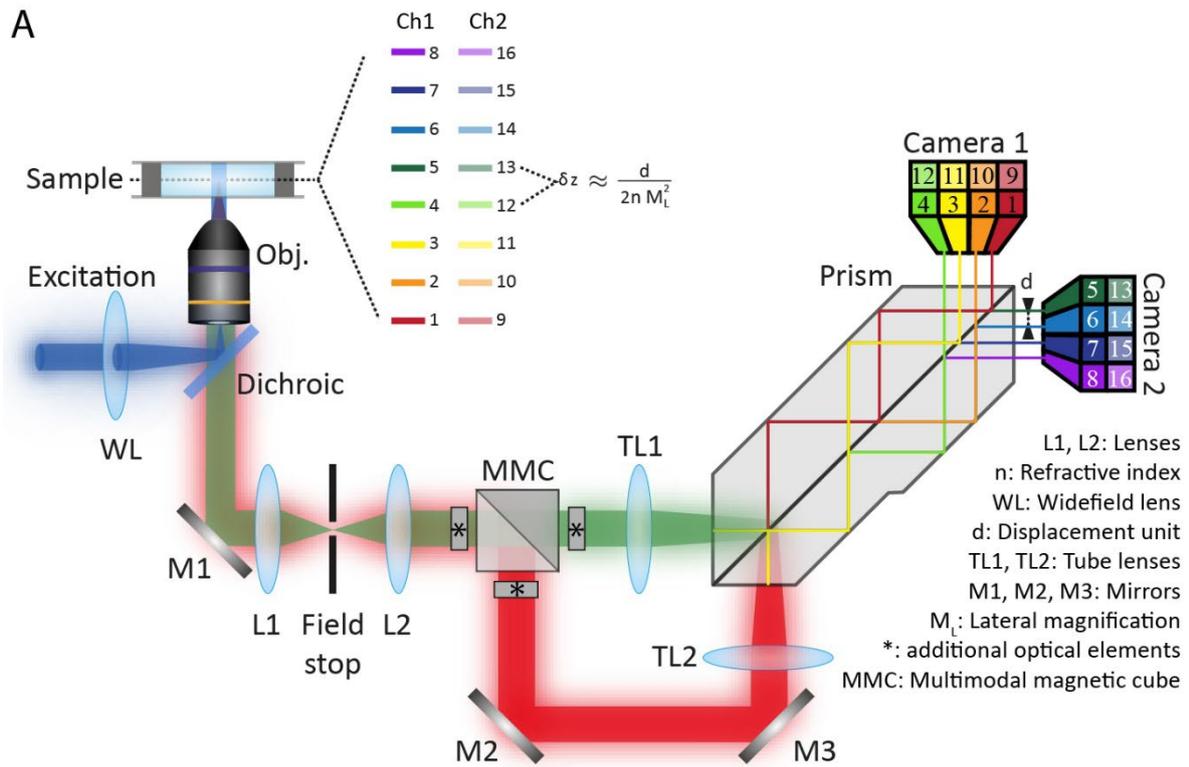

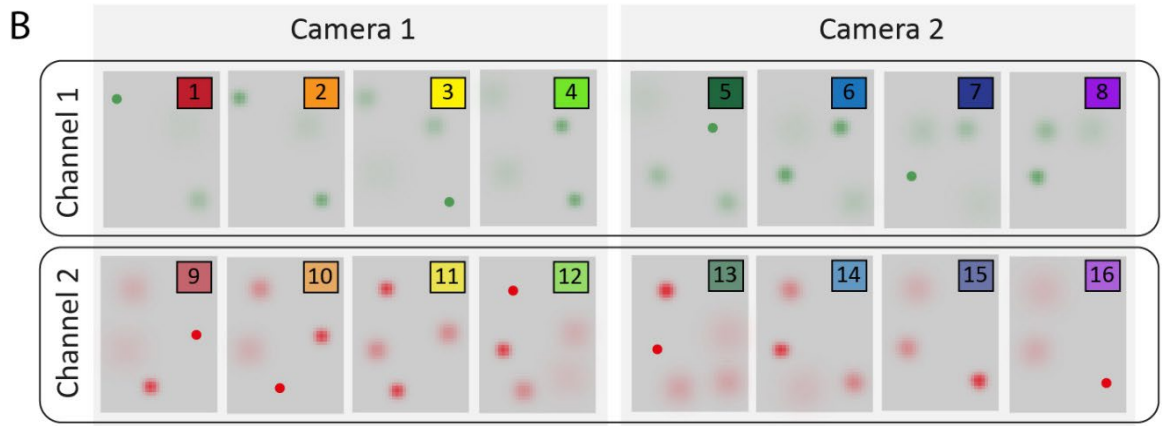

*Figure 1: Optical layout of the custom-built M³Scope. (A) Emission collected by the objective is relayed through an 8f detection system containing a multimode magnetic cube (MMC) that enables insertion of dichroic mirrors or beamsplitters to create two independent detection arms. Mirrors (M2, M3) direct the secondary beam into the multiplane prism from a different face, generating two channels (Ch1, Ch2). The prism projects eight focal planes per channel onto two CMOS cameras, resulting in simultaneous 16-plane acquisition. Positions marked * allow insertion of optical elements. (B) Schematic of plane projections from both channels onto the two cameras, showing eight depth-resolved planes per channel.*

This design provides several positions where additional optical elements can be inserted (indicated by * in **Fig. 1A**). Components placed before the MMC affect both detection channels simultaneously, while those positioned after the MMC act only on their respective channels. This modular configuration allows optical components to be added flexibly depending on experimental requirements. For example, spectral filters can be inserted to isolate specific emission bands for spectral cleanup. Additionally, full-aperture obstruction targets can be placed in the infinity plane to implement Fourier filtering. In this configuration, blocking the central portion of the pupil suppresses the undiffracted (zero-order) light, enhancing scattering contrast and producing a darkfield-like imaging mode without the need for a traditional darkfield condenser. [34] Similar aperture masks can also be used to suppress stray light or to facilitate phase imaging strategies. Because all beams remain collimated in the infinity space, inserting or removing these additional elements does not disturb focus or alignment, ensuring robust operation when switching modalities. Once the mirrors M2, M3, and the tube lens TL2 are aligned, only minimal realignment is required when changing modalities.

Switching between modes can be achieved in less than five minutes simply by exchanging the dichroic or beamsplitter in the MMC and, if needed, adding or removing additional elements at the designated positions. This flexibility enables a wide range of multimode configurations. Inserting a dichroic mirror creates a dual-color fluorescence mode, allowing different dyes to be imaged simultaneously. Replacing it with a 50:50 polarizing beamsplitter enables polarization-based imaging, capturing s- and p-polarization intensities separately. Finally, correlative multimodal imaging, such as combining brightfield, fluorescence, phase, darkfield, or backscattering modes, can be achieved by inserting a beamsplitter together with appropriate spectral elements in each channel. This straightforward architecture integrates fast imaging, volumetric multiplane acquisition, and flexible multimode operation within a single, alignment-stable system. This enables rapid, multi-channel access to structural and dynamic information, without complex optical realignment or redesign.

A calibration of the multiplane geometry and potential shifts between the two channels is however needed. This has been performed using a *z*-stack of fluorescent multicolor nanoparticles, each containing both green and red dyes, drop-casted on glass. The procedure, detailed in **SI.1**, followed the method reported in our previous work [10] and has been extended to ensure spatial alignment between the two imaging channels. Corresponding multiplane images from both channels have been cross-correlated to calculate relative lateral shifts, after which the regions of interest (ROIs) in the second channel have been adjusted to match those in the first one. For tracking of single emitters, an additional super-resolution calibration can be done for sub-pixel alignment corrections between the two channels, as described below.

*Dual-color imaging for multiscale dynamics*

The first imaging mode implemented in the system is dual-color fluorescence, where the two emission paths capture distinct spectral ranges. The emission was split by a 532 nm dichroic beamsplitter to separate the emissions from green and red fluorophores, excited at 488 nm and 561 nm, respectively. For spectral cleanup, a 595/40 nm bandpass filter has been placed in channel 1 and a 525/50 nm bandpass filter in channel 2, while a 488/561 nm multiband notch filter is positioned before the cube blocked excitation light (**Fig. 2A**).

To verify alignment and focus matching between both detection channels, 200 nm fluorescent polystyrene nanoparticles containing both green and red dyes were spin-cast onto a glass coverslip and imaged simultaneously in both channels. **Figure 2B** shows representative images of a single particle across all eight planes in both channels, confirming coincident focal depth and lateral localization. Particle localizations have been matched between channels, demonstrating consistent overlap across the entire field of view.

In addition to the multiplane calibration described in the previous section, which aligns the field of view and focal distance of all planes, a super-resolution calibration has been performed to quantify and correct residual offsets between the two channels: the positions of corresponding emitters in both channels have been localized with sub-pixel precision, and the resulting coordinate pairs have been used to calculate and correct for residual translational, rotational, and scaling offsets in the second channel. For a dataset acquired with a 250 ms exposure time, 7551 paired localizations could be identified. After calibration, the mean displacement between corresponding localizations in the two channels is zero, with residual standard deviations of 39.1 nm in *x* and *y* directions, and 65.3 nm in the *z* direction.

To assess whether these residuals originated from measurement noise or optical misalignment, time-lapse images of stationary particles were analyzed. The localization precision was 13 nm and 14 nm in *x* and *y*, and 23 nm in *z*, indicating that the observed residuals mainly reflect the combined localization accuracy of both channels rather than true misalignment. Of note, the values in *x* and *y* direction are below the effective pixel size (95nm), and the colocalization error in the axial direction is significantly lower than the spacing between adjacent planes (580nm), confirming sub-pixel registration precision across all dimensions.

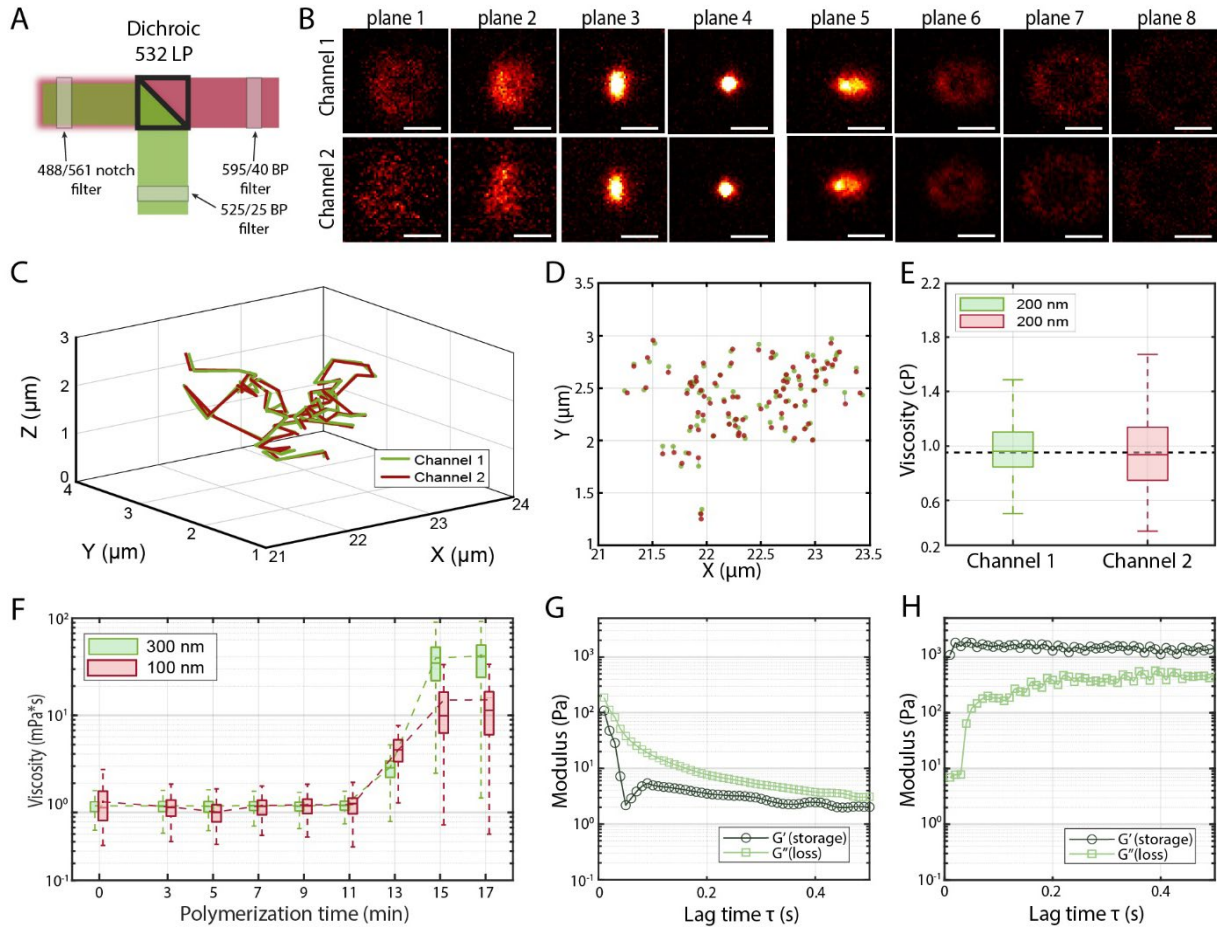

*Figure 2: Fast 3D dual color fluorescence imaging. (A) Insertion of a 532 nm LP beamsplitter into the optical cube enabled dual-color imaging with extra filters for spectral clean-up. (B) Images of a multicolour particle over different planes in both channels. Images in different planes originate from different focal planes. The two channels capture different spectral ranges and different dyes from the particles. Scalebars 2 μm. (C) Overlapping particle trajectories from both channels after super-resolution calibration. (D) XY projection of the localisations of the trace in panel C. (E) Viscosity measurements from the trajectories of 200 nm multicolour fluorescent polystyrene nanoparticles. The black dashed line indicates the reference value of 0.934 mPa*s for the viscosity of water. (E) Viscosity trend during polymerization, highlighting the divergence between particle sizes after gelation. (G) Average loss- and storage moduli as calculated from the Brownian motion of 300 nm particles, at 5 min into polymerisation. (H) Average loss- and storage moduli as calculated from the Brownian motion of 300 nm particles, at 15 min into polymerisation.*

To verify that both detection channels provide consistent information under dynamic conditions, the same dual-labelled 200 nm fluorescent nanoparticles have been dispersed in water, and 100 fps time-lapse images have been acquired with an exposure time of 10 ms per frame. Representative trajectories from one particle in both detection channels are shown in Figure 1C, with the corresponding positional overlap in Figure 1D, confirming good agreement between the two channels. For nanoparticles diffusing in water, the inter-channel colocalization error increased to 85.8 nm in *x* and *y*, and 178 nm in *z*. The reduced precision compared to the static sample primarily reflects (i) lower signal-to-noise ratio at shorter exposure times and (ii) particle motion during image integration time.

To quantify the particle dynamics, diffusion coefficients were calculated from the trajectories detected in both channels. To ensure reliable statistics, only trajectories containing more than 20 data points have been included in the analysis. Channel 1 exhibited higher detection efficiency, yielding fewer but longer trajectories (747 tracks, 54 points on average) than channel 2 (920 tracks, 29 points on average).

For each trajectory, the mean-squared displacement (MSD) has been constructed, and the diffusion coefficient has been determined from the slope of a linear fit to the first four MSD points [35]. Using the Stokes-Einstein equation for viscous liquids, the apparent water viscosities could be calculated (**SI.**

**2A**). The resulting values are 0.99 ± 0.43 mPa*s (mean ± std) for channel 1 and 0.95 ± 0.77 mPa*s for channel 2, both in good agreement with the reference viscosity of water at 23 °C (0.94 mPa*s, **Fig. 1D**). The higher spread in channel 2 arises from the shorter trajectories and lower signal from the dye, which increases the uncertainty in the first datapoints of the MSD. Despite these differences in signal and track length, both channels yielded consistent and quantitatively accurate measurements of particle diffusion in water.

This level of precision in capturing Brownian dynamics demonstrates the system's suitability for quantitative dynamic measurements. Having validated that both detection channels provide consistent diffusion coefficients, we next applied the setup to investigate the local viscoelastic properties of complex media using dual-color passive microrheology (see **SI 1A** for details). This approach enables simultaneous tracking of tracer particles of different sizes, providing direct insight into how probing length scales influences the measured polymer dynamics.

Particle tracking has been performed in a polymerizing a polyacrylamide (PAA) network, with two different carboxylate-modified polystyrene nanoparticles as probes: 100 nm diameter particles containing a red dye (561 nm excitation) and 300 nm diameter particles with a green dye (488 nm excitation, **Materials and Methods**). The particle motion has been tracked simultaneously in both channels. During the initial 11 minutes of polymerization, the network remained predominantly liquid, with comparable viscosities for the two probe sizes (mean ± std): 1.17 ± 0.22 mPa*s for the 100 nm particles and 1.18 ± 0.38 mPa*s for the 300 nm particles (**Fig. 2F**). These similar values indicate minimal hindrance of either probe, consistent with a homogeneous fluid behavior at the scales probed.

After the gelation point (~13 minutes), particle dynamics began to diverge. The 100 nm particles probed a viscosity of 14.5 ± 12.1 mPa*s at 17 minutes, whereas the 300 nm probes experienced a substantially higher viscosity, reaching 40.8 ± 23.3 mPa*s. The pronounced variance increase, from 19-32% before gelation to 57-84% after, indicates the emergence of structural heterogeneity as the polymer network forms, consistent with kinetic Monte Carlo modeling results at high yield. [36]

This size-dependent disparity and increased variability mark the breakdown of the classical Stokes-Einstein relation, where the assumption of purely viscous behavior no longer holds [37,38]. While the calculated viscosities suggest scale-dependent dynamics, their absolute values are underestimated because the classical Stokes-Einstein equation assumes purely viscous behavior, therefore overlooking the elastic contribution. To fully describe the viscoelastic behaviour, we next applied the generalized Stokes-Einstein relation to determine the frequency-dependent storage and loss moduli from the mean-squared displacement.

This analysis (**SI 1C**) revealed that up to 13 minutes, the loss modulus exceeded the storage modulus, indicating liquid-like behavior (**Fig. 2G** and **SI 2A**). Beyond this point, a crossover between the moduli signals marked the gelation point. At later times, the 300 nm particles reported higher stiffness values (up to 2 kPa storage modulus) compared to the 100 nm particles, which saturated at ~0.5 kPa (**Fig. 2H** and **SI 2C**). Both viscosity and stiffness reached a plateau after ~15 minutes of polymerization (**SI 1C**), likely because particle displacements were below the detection limit rather than a cessation of polymer growth. This limitation leads to an underestimation of rheological parameters at high polymer or crosslinking densities which is addressed in the following section using rotational tracking.

Overall, these results demonstrate that dual-color multiplexed tracking combined with generalized microrheological analysis, captures heterogeneous and scale-dependent viscoelastic behavior that would remain hidden in single-particle or single-channel measurements.

*Polarization imaging for rotational tracking*

As discussed in the previous section, while translational particle tracking provides valuable insights into microrheological behavior, it has an intrinsic upper limit in the viscosity range it can accurately probe. In the M$^3$Scope, the localization precision of immobilized particles is 13-14 nm in the *x* and *y* directions, and 23 nm in *z*, corresponding to a minimum detectable diffusion coefficient of 0.016 μm$^2$/s (at a frame rate of 4 Hz). When Brownian displacements fall below this threshold, as in highly viscous or stiff materials, the apparent motion becomes dominated by localization noise, leading to overestimated diffusion coefficients and, consequently, underestimated viscosities. This limit corresponds to viscosities above ~130 mPa*s. In imaging modes with lower signal-to-noise ratios, the localization precision decreases further, reducing the maximum measurable viscosity range.

To illustrate this limitation, we tracked gold bipyramids (AuBPs, **Fig. 3B**) diffusing in glycerol–water mixtures of increasing viscosity. To visualize the nanoparticles, obstruction targets have been placed in the Fourier plane of both channels to block the undiffracted light, generating a dark-field configuration that isolates the scattered signal from the AuBPs (**Fig. 3A**). Translational diffusion coefficients have been converted to apparent viscosities using the Broersma slender-body correction for anisotropic nanoparticles on the Stokes–Einstein relation (**SI.1A**). The calculated viscosities 26.08 ± 18.02 mPa*s, 39.35 ± 27.24 mPa*s, 55.58 ± 44.71 mPa*s, and 57.36 ± 45.88 mPa*s for mixtures of 80%, 85%, 90%, and 95% (v/v) glycerol, respectively, were all substantially lower than the reference values of 75, 130, 241, and 484 mPa*s (**Fig. 3K**). Even at 75 mPa*s, translational motion could not be reliably resolved, as the low signal-to-noise ratio of the dark-field signal led to large localization errors. However, although translational motion becomes undetectable in dense or highly viscous media, nanoparticles can still undergo rotational diffusion within their local confinement. [38]

Polarization-resolved imaging has been implemented in the M$^3$Scope to measure 3D rotational dynamics of the bipyramids at high speed. A 50/50 polarizing beamsplitter was inserted in the multimode cube, and custom obstruction targets have been placed in both detection channels (**Fig. 3A**). The beamsplitter separated the scattered light into s- and p-polarized components while the obstruction targets enabled a dark-field configuration that suppressed background light and isolated the nanoparticles' forward-scattered signal.

Owing to their anisotropic shape and scattering response, gold bipyramids are ideally suited for monitoring rotational dynamics (**Fig.3B**). Indeed, under unpolarized illumination, AuBPs exhibit their strongest plasmon resonance along the long axis (longitudinal plasmon resonance) [40], making the scattered intensity highly orientation-dependent. By splitting the scattered light into s- and p-polarized components, changes in the in-plane orientation (θ angle) of the particles can be monitored through the relative intensity difference between both channels, while variations in the total scattered intensity provide information on the out-of-plane angle (*φ*). Combined with multiplane detection, this configuration enables fast, 3D tracking rotational dynamics in individual nanoparticles.

To validate the accuracy of in-plane rotation tracking, a 2D experiment has been performed using AuBPs spin-coated on glass (see **Materials and Methods**). Because the particles were immobilized, the effect of particle orientation on the detected signals have been evaluated by rotating, at a defined angular speed, a λ/2 waveplate inserted before the multimode cube. In this configuration, rotating the waveplate changes the polarization orientation of the scattered light reaching the detector, producing the same intensity fluctuations that would occur from real in-plane rotation. Two datasets have been acquired: one with 10 ms exposure time with a 25 °/s rotation speed, and another at 100 ms exposure time with a 5 °/s rotation speed. The recorded s- and p-polarized intensities exhibited sinusoidal modulation with the expected π/2 phase shift between the two channels (**Fig. 3E**), while the total scattering intensity remained constant. Autocorrelation analysis of the intensity traces (**see Materials and Methods**) enabled the retrieval of the periodic time $T_{period}$ corresponding to a full 360° rotation

(**Fig. 3F**). The calculated measured angular velocities, 28.66 ± 8.78 °/s and 5.13 ± 0.13 °/s for rotation rates of 25 and 5 °/s, respectively (mean ± std; **Fig. 3G**), closely match the imposed values. These results confirm that polarization-resolved dual-channel imaging accurately quantifies the in-plane rotational dynamics (θ) of anisotropic nanoparticles.

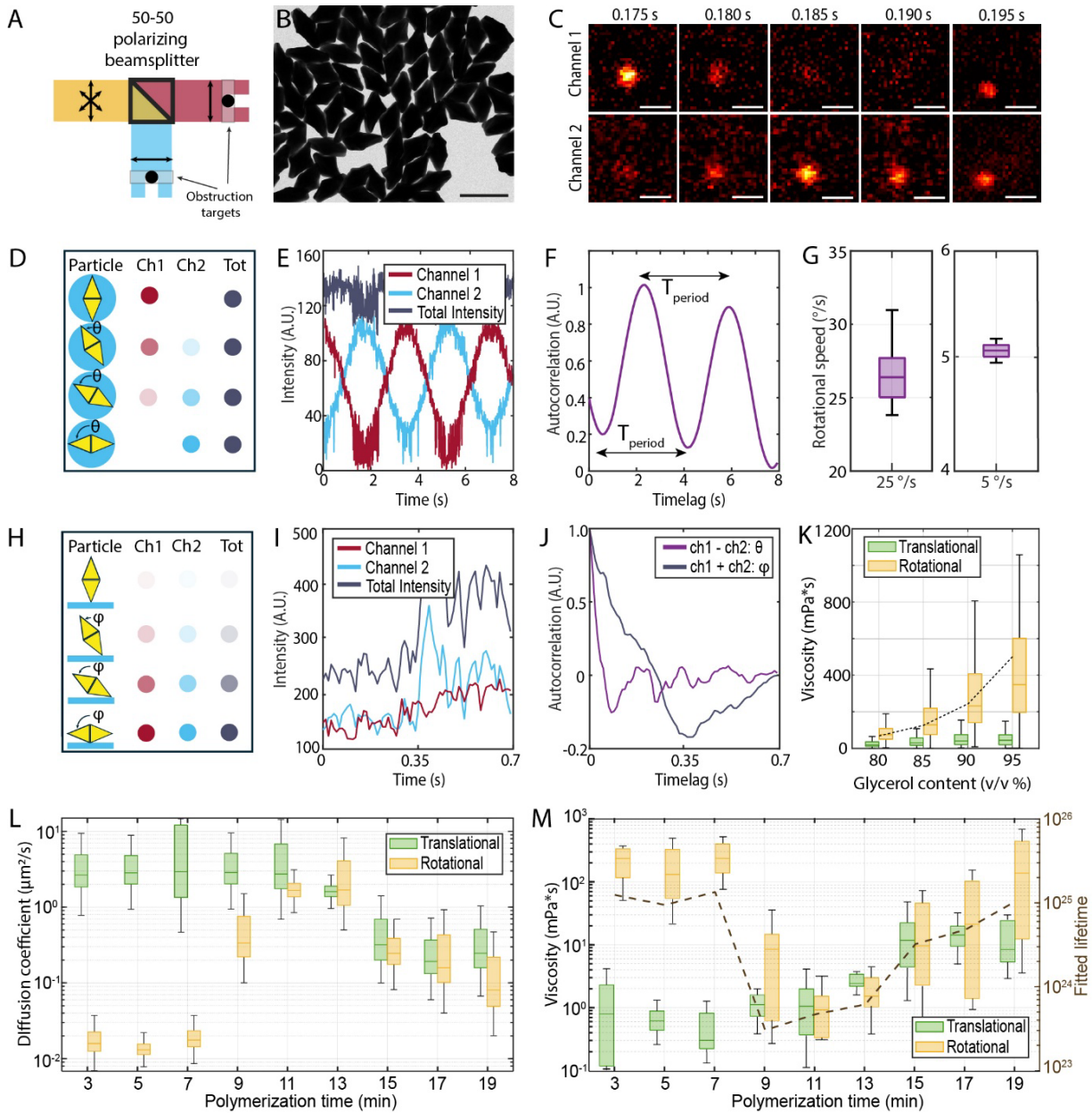

*Figure 3: Instant 3D rotational tracking by polarization imaging. (A) Insertion of a polarising beamsplitter into the optical cube separates s- and p-polarisation components. Obstruction targets in every channel to enable dark-field-like mode. (B) TEM image of 182 nm x 91 nm gold bipyramids. Scalebar 200 nm. (C) Particle scattering intensities fluctuate over time, indicating rotational motion. Scalebars 2 μm. (D) Rotation in the in-plane angle theta leads to a different ratio of intensities between the two channels. (E) Sample intensity trace of a single particle during 2D measurement: channels 1 and 2 exhibit sinusoidal modulation with a π/2 phase shift, while the total intensity remains constant. (F) Autocorrelation function of a single particle during 2D measurement reveals periodic behaviour from which the rotational speed of the waveplate can be recalculated. (G) Recovery of the waveplate's angular speed, shown here for 25°/s and 5°/s datasets. (H) Total intensity encodes the out-of-plane angle (φ), enabling full 3D rotational tracking. (I) Intensity traces during particle diffusion in a glycerol 80% mixture. (J) Autocorrelation functions for total intensity (φ) and difference (θ). (K) Rotational and translational measurements of viscosity in water–glycerol mixtures; the dashed line indicates ground truth viscosity. (L) Trend of diffusion coefficient during polyacrylamide polymerisation, starting with an underestimation of the rotational diffusion coefficient at low viscosities, while the rotational diffusion coefficient becomes more reliable from 11 minutes into the polymerisation. (M) Viscosity trend during polyacrylamide polymerisation, showing that rotational microrheology extends to higher viscosities compared to translational tracking. The dashed brown line depicts the average exponential time coefficient.*

Beyond in-plane rotation, φ can also be extracted from polarization-resolved imaging. Variations in φ determine how the particle's long axis projects onto the focal plane, thereby modulating the total scattered intensity collected across both detection channels (**Fig. 3H**). Dual-channel detection is essential for measuring the 3D rotation of particles, as it enables simultaneous acquisition of both orthogonal polarization components and thus provides access to the complete scattered intensity. When particles move freely in three dimensions, however, intensity fluctuations can also arise from axial motion as they move in and out of focus. To disentangle these effects, multiplane acquisition in the M³Scope captures the full intensity distribution of each particle across several focal planes simultaneously. This ensures that changes in total intensity are correctly attributed to rotational dynamics rather than z-displacement and enables robust 3D localization of moving particles.

By combining the information from the in-plane angle ($\theta$) and out-of-plane angle $\varphi$ across the different planes, the full 3D rotational trajectories can be reconstructed. From these, the rotational diffusion coefficient can be determined, and the local viscosity can be calculated by the Stokes-Debye-Einstein equation [38] (**SI**). Of note, because particle positions are tracked simultaneously, translational and rotational motion can be quantified in parallel, providing a complete description of nanoparticle dynamics and enabling direct correlation between translational and rotational diffusion.

To demonstrate this capability, we also measured the viscosity of water–glycerol mixtures by tracking the 3D rotational diffusion of AuBPs. For each trajectory, intensity time traces have been extracted from channel 1, channel 2, and the combined total signal (**Fig. 3I**). Autocorrelation analysis of the total intensity yield $\varphi$, while the analysis of intensity difference between channels provides $\theta$(**Fig. 3J**). Measured viscosities for glycerol concentrations of 80, 85, 90, and 95% (v/v) were 86 ± 58 mPa*s, 162 ± 134 mPa*s, 350 ± 391 mPa*s, and 440 ± 391 mPa*s, respectively. These values that closely match the theoretical viscosities of 75, 130, 241, and 454 mPa*s (**Fig. 3L**).

In contrast to translational tracking, which consistently underestimated viscosities, rotational tracking thus remained accurate and stable throughout this range. These results demonstrate that polarization-resolved multiplane imaging in the M³Scope uniquely enables quantitative 3D rotational tracking into the high-viscosity regime inaccessible to conventional translational approaches.

Having validated rotational tracking in highly viscous glycerol mixtures, we applied this approach to a more complex viscoelastic system. In the polyacrylamide (PAA) polymerization experiments described in the previous section, translational tracking failed at late stages as particle displacements fell below the localization limit. Here, we tested whether rotational tracking could overcome this limitation and provide a more complete picture of the network's mechanical evolution. The same polymerization system has been used, with AuBPs dispersed in the matrix. During gelation, polarization-resolved intensities have been continuously recorded, and translational and rotational viscosities could be extracted at successive time points (**Fig. 2G**).

In agreement with the previous results, translational tracking provided accurate viscosity values before the gelation point. Up to ~9 min, the viscosities determined from translational motion ranged between 0.56 and 1.23 mPa*s, consistent with the values obtained from tracking fluorescent beads (see previous section). In this regime, rotational tracking overestimated viscosities because the bipyramids rotated multiple times within the exposure period, averaging out the polarization signal and reducing the sensitivity to rotational diffusion.

Between 11 and 17 minutes, both rotational and translational tracking yielded comparable viscosities, reflecting consistent measurements of the progressively slowing dynamics. Rotational tracking yielded 1.97 ± 0.59, 1.84 ± 1.55, 13.28 ± 8.22, and 16.72 ± 15.36 mPa*s, whereas translational tracking gave 1.25 ± 1.04, 2.06 ± 0.50, 10.35 ± 8.07, and 16.96 ± 11.10 mPa*s at the corresponding time points. This convergence indicates that particle rotation had slowed sufficiently to be resolved within the exposure period.

To determine the point at which rotational tracking became reliable, we analyzed the autocorrelation of the rotational intensity fluctuations in channel 1. Rapid rotation resulted in minimal intensity variation within a frame, yielding autocorrelation curves with long lifetimes. As the polymer network formed and rotation slowed down, the correlation lifetime decreased. As shown in **Fig. 3M**, a clear drop in lifetime occurred around 9 minutes, indicating that beyond this point the rotational signal could be captured reliably.

At later stages of polymerization (≥19 minutes), when the network became dense and translational motion was largely arrested, the two methods diverged. Translational tracking plateaued around 13.45 ± 10.83 mPa*s, as particles became translationally immobilized, whereas rotational tracking continued to report increasing resistance, reaching 40.62 ± 38.99 mPa*s. This behavior reflects the regime where translational displacements fall below the localization limit, while rotational motion, though slowed, remains measurable, preserving sensitivity to the increasing mechanical resistance of the medium. Together, these results highlight that rotational tracking extends the measurable viscosity range and remains sensitive even when translational motion becomes undetectable.

*Multimodal imaging for probing structural heterogeneity*

In addition to dual-color and polarization imaging, the modular design of the $M^3$Scope enables simultaneous imaging with two different contrast mechanisms within a single optical setup. With the appropriate dichroic beamsplitter into the multimode cube and filter on the lamp, different imaging modalities can be combined, including fluorescence with brightfield, dark-field, phase, or polarization-based contrast. This versatility allows users to correlate structural, chemical, and dynamic information in real time, without modifying the core optical alignment. Such multimodal acquisition is particularly valuable in soft-matter and biological systems, where label-free contrasts reveal morphology or refractive changes, while fluorescence tracks molecular or dynamic processes.

Here, we demonstrate this capability by probing structural heterogeneity in pNIPAM networks during a temperature-induced phase transition. pNIPAM exhibits a strong temperature-dependent refractive index: at low temperatures, the polymer adopts an open-chain, water-soluble state with a refractive index of 1.366, which increases linearly as the chains collapse into the closed-chain phase at elevated temperatures [41]. These refractive-index changes provide intrinsic contrast, allowing structural transformations to be directly monitored through bright-field imaging.

To link these structural variations with local dynamics, the $M^3$Scope was adapted to combine bright-field and fluorescence modalities. A 532 nm beamsplitter was placed in the multimode cube to divide the emitted light into two detection paths. In the brightfield channel, a 655 nm long-pass filter was used for spectral cleanup, and another 655 nm long-pass filter was mounted on the illumination lamp. To monitor viscoelastic properties, embedded 200 nm fluorescent carboxylate-modified polystyrene nanoparticles have been tracked in the fluorescence channel using 488 nm excitation and a 525/50 nm emission filter (**Fig. 4A**; **Materials and Method**). During image acquisition, the temperature gradually increased from 30 to 35 °C, crossing the phase transition point at 32 °C, and subsequently cooled down to room temperature.

During heating, the brightfield channel revealed structural changes associated with the pNIPAM phase transition. At 30 °C, the bright-field signal from the polymer network appeared homogeneous, consistent with the water-soluble state (average intensity: 131.01 ± 9.1 counts/px). As the temperature approached the phase transition point (~32 °C), the image texture evolved into coarser, grainier intensity variations indicative of polymer chain collapse and local refractive index fluctuations. The average intensity decreased to 111.78 ± 10. counts/px at 35°C (**Fig. 4.B**). To quantify these changes, the gradient magnitude of the intensity was calculated as a measure of image texture. With the appearance of a grainier structure, the local pixel-to-pixel differences decreased, and the gradient dropped from

39.65 ± 4.1 counts at 30°C to 27.0 ± 7.5 counts at 35°C (**Fig. 4.C**), indicating partial homogenization as collapsed domains formed.

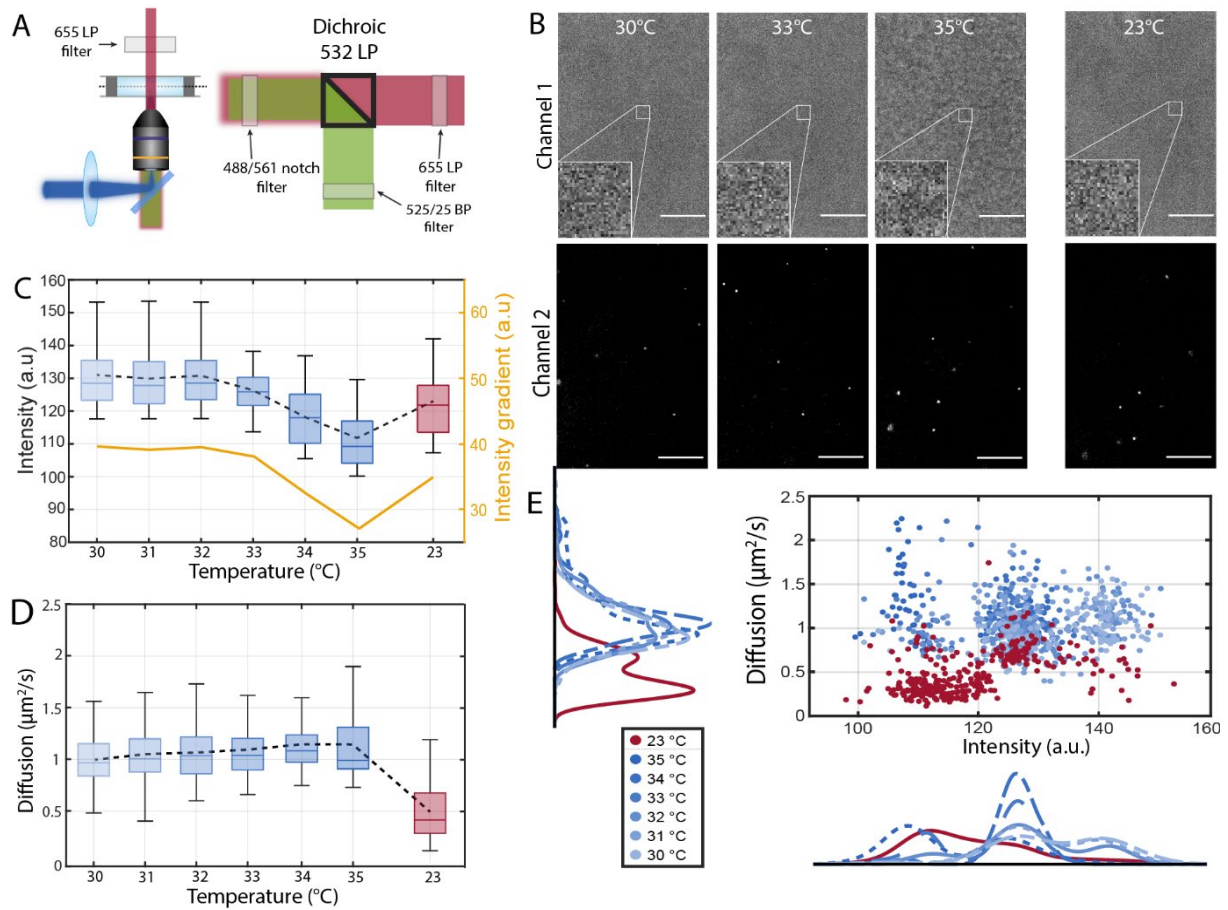

*Figure 4: Multimodal imaging for simultaneous tracking of diffusion dynamics and structural heterogeneity in a phase-separating pNIPAM network. (A) Insertion of a color beamsplitter into the optical cube enables spectral separation between imaging modalities. Additional filters were added for spectral clearance of both brightfield emission and detection. (B) Raw intensity images over time show that above the 32 °C phase transition temperature of pNIPAM, intensity graininess increases while overall intensity decreases; both effects are reversible upon cooling. Scalebars 10 μm. (C) Intensity and intensity gradient trend during heating and recovery of the pNIPAM–water mixture. (D) Corresponding diffusion coefficient trend, with higher nanoparticle mobility observed at elevated temperatures. (E) Correlation of particle diffusion and brightfield intensity during the phase transition. While diffusion is primarily temperature-controlled (left histogram), intensity exhibits three co-existing states: a bright state at low temperatures, an intermediate state, and a dark state dominating above the transition.*

To connect structural changes in pNIPAM to dynamics, nanoparticle dynamics have been simultaneously tracked in 3D using the fluorescence channel. Particle mobility increased with temperature: the diffusion coefficient rose from 0.987 ± 0.217 μm²/s at 30 °C to 1.135 ± 0.361 μm²/s at 35 °C, corresponding to a decrease in local viscosity from 1.155 ± 0.271 mPa*s to 1.033 ± 0.257 mPa*s (**Fig. 4.D**). This indicates that the temperature-dependent increase in thermal motion dominates over the phase effect: despite that the collapse of pNIPAM chains into the closed-chain conformation could locally trap particles and increase viscosity, the enhanced thermal energy at higher temperatures leads to faster particle diffusion and an overall decrease in measured viscosity.

After heating, the sample was cooled back down to 23 °C and a final measurement has been recorded 3 minutes into the cooling down. In the tracked particle dynamics, diffusion coefficients largely returned to low-temperature values. For the brightfield intensity, both the average intensity and graininess became higher than at 35 °C but not yet on the initial value. Apparent graininess was largely gone, suggesting the network had mostly regained its homogeneity.

Correlative analysis between the two modalities has been performed by extracting the brightfield intensity from channel 1 at the spatial coordinates of the tracked particles in channel 2, thereby assigning each localization a corresponding structural intensity value (**Fig. 4E**). The temperature-dependent evolution of the intensity distribution revealed distinct regimes of structural heterogeneity. At 30 °C, particle-associated intensities displayed a bimodal distribution, indicating the coexistence of bright and dim regions that correspond to the initial stages of polymer domain formation. As the temperature increased, the lower-intensity state progressively dominated, reaching its maximum prevalence at 34 °C. At 35 °C, a third, even darker population emerged, marking the formation of highly collapsed polymer regions. During this transition, particle diffusion showed small overall increases driven by temperature, but with large variance due to local structural heterogeneity within the polymer network. Upon cooling back to 23 °C, the diffusion coefficients decreased again, yet the corresponding intensity histogram exhibited a right-tailed distribution - evidence that the structural relaxation and homogenization of the network were still ongoing.

These results demonstrate how multimodal imaging provides a direct link between structural (molecular) evolution and local particle dynamics, capturing both the reversible and heterogeneous nature of the pNIPAM phase transition. Importantly, the quantitative phase imaging channel operates without additional calibration, as it leverages the combined multiplane phase-retrieval approach [18], enabling direct extraction of absolute phase information.

# Conclusions

We have demonstrated the versatility of the M³Scope, a microscope that enables fast 3D imaging (>100 fps) in multiple mode simultaneously, with flexible and alignment-stable modality switching, validated through a series of microrheology experiments. M³Scope bridges the gap between speed and multi-mode in 3D imaging. Indeed, traditional 3D imaging techniques each face trade-offs: confocal microscopy offers high axial resolution but is slow due to point-by-point scanning, light-sheet microscopy improves speed but relies on sequential acquisition or complex illumination schemes [42,43], and conventional multiplane systems achieve millisecond acquisition yet rely on grating-based splitting or angled coupling into the prism [16,25,26], which demand precise alignment and lack flexibility. The M³Scope overcomes these limitations with a simple design based on a modular beamsplitter cube.

Beyond polymer physics demonstrated here, these multimode configurations open broad opportunities across soft-matter and biological research. Dual-color imaging can monitor nanoparticle transport and subcellular diffusion, polarization-resolved tracking enables quantification of local forces and torques in anisotropic probes, and multimodal imaging provides a label-free route to correlate structural reorganization with molecular motion in processes such as cytoskeletal remodeling, protein condensate formation, or cell-matrix mechanotransduction. [44,45]. Importantly, the same concept can be easily adapted to other contrast mechanisms, such as quantitative phase, darkfield, or brightfield imaging, without substantial modification of the optical layout. With its multimode imaging and its unprecedented flexibility, M³Scope provides a general platform for studying non-equilibrium, heterogeneous, and dynamic systems across physics, materials science, and biology.

# Materials and Methods

*Microscope*

Excitation was provided by multiple laser lines depending on the experiment, all focused at the back focal plane of a 60× oil-immersion objective (Olympus, UPlanSApo60XOil, 1.4 NA) via a wide-field lens (Thorlabs, plano-convex, f = 150 mm): 488 nm laser line (100 mW, SpectraPhysics), 561 nm laser line (Coherent). The excitation light was guided into the sample using a dichroic mirror reflecting 488, 561, and 640 nm (Chroma Technology). Emission collected by the objective was collimated and passed through a telecentric 4f lens system (L1: Thorlabs, plano-convex, f = 140 mm; field stop; L2: Thorlabs, plano-convex, f = 140 mm; tube lens: Thorlabs, plano-convex, f = 200 mm), where L1 formed an intermediate image cropped by the field stop and the tube lens produced the final image. A proprietary prism (patent EP3049859A1; LOB-EPFL, Lausanne; Schott SA, Yverdon) split the light into eight beams with different path lengths, enabling multiplane imaging. [10] The imaging planes were collected by two CMOS cameras (four planes per camera). Fast synchronized acquisition was achieved by using the output trigger of camera 1 as input trigger for camera 2, while focus was maintained by moving the sample stage with three uni-axis PI motors (E-871-1A1) controlled via Micro-Manager. [46,47]

*2-channel plane calibration*

Image planes were calibrated using 200 nm fluorescent polystyrene nanoparticles spin-coated on glass. A 50:50 beamsplitter projected 16 planes onto the two cameras. Multiple z-stacks were acquired over a 10 μm range in 100 nm steps. For each z-step, all planes were imaged sequentially on both cameras to sample overlapping depths for alignment. Plane positions were initially determined by intensity thresholding to define each plane's ROI.

Within each channel, relative zoom, translational shifts, and rotational misalignments between planes were corrected using the *imregcorr* function in MATLAB. Corrections between corresponding planes of the two channels were then determined using a similar approach. x–y alignment across planes was achieved via pixel-wise correlation, and axial distances were determined by Gaussian fitting of intensity gradients, allowing normalization of intensity across depths (Fig. SI.A).

After this basic calibration, a super-resolution alignment was applied, where particles were first localized in x and y in each plane. Within each channel, coordinates across planes were compared and small translation and rotation corrections were calculated and stored. After applying intra-channel corrections, corresponding planes across channels were compared, and the final transformation between channels was calculated using *Procrustes* function in MATLAB. Both corrections (intra-channel and inter-channel) are applied sequentially during particle detection, ensuring sub-micron registration and accurate trajectory matching across the two channels.

*Sample: Polyacrylamide network*

Hydrogels were prepared from 4 w/v% acrylamide, 0.17 w/v% N,N'-methylenebisacrylamide, 0.25 w/v% ammonium persulfate (APS), and 0.25 v/v% N,N,N',N'-Tetramethylethylenediamine (all purchased from Sigma Aldrich). Polymerization was initiated by adding APS (time point 0). The mixture was prepared at double concentration and diluted with aqueous solutions containing fluorescent polystyrene nanoparticles (ThermoFisher, 1:5000) or lab-synthesized gold bipyramids (sonicated 10 min). 10 μL of the mixture was drop casted onto 1-hour ozonated coverslips, and the hydrogel was shielded from air with a 120 μm imaging spacer (9 mm diameter, Secure Biolabs) and a second ozonated coverslip.

*Sample: Multicolour nanoparticles in water*

TetraSpeck™ microspheres (0.2 µm, blue/green/orange/dark red; ThermoFisher) were diluted 1:5000 in Milli-Q water. 10 µL of the suspension was placed between two ozonated coverslips with a 120 µm, 9 mm diameter imaging spacer (Secure Biolabs).

### Synthesis of gold bipyramids

Gold bipyramids were prepared following a recently reported 3-step protocol (**2024 Bevilacqua**).

In a first step, pentatwinned decahedral seeds were prepared by fast reduction of $HAuCl_4$ with $NaBH_4$ in the presence of citric acid and hexadecyltrimethylammonium chloride (CTAC) (**2016 Sanchez-Iglesias**). In a typical synthesis, a solution containing 50 mM CTAC, 5 mM Citric acid and 0.25 mM $Au^{3+}$ was prepared and left at room temperature for 15 min to ensure homogeneity. After this time, 250 µL of ice-cold 25 mM $NaBH_4$ was added under vigorous stirring (900 – 1000 rpm), after which the solution should turned color from light yellow to light brown. After 10 min to ensure borohydride decomposition, the vial was placed in a bath at 80 ºC for 90 min. After heat treatment, the vial color turned from light brown to bright red, showing a characteristic extinction band around 529 nm.

In the second step the decahedral seeds were overgrown into gold nanorods. A solution containing 100 mL of 8 mM hexadecyltrimethylammonium bromide (CTAB) and 0.125 mM $Au^{3+}$ was stirred and kept at 20 ºC. After 20 min to ensure temperature equilibration, 250 µL of 100 mM ascorbic acid was added to the solution under vigorous shaking. After the solution turned colorless, 350 µL of the previously prepared decahedral seeds was injected under vigorous shaking. The reaction mixture was then stored at 20 ºC overnight. The particles were centrifuged and purified using depletion-induced precipitation. The dimensions of the obtained pentatwinned nanorods were 85 nm × 23 nm.

Finally, bipyramids were obtained by overgrowth of the nanorods in the presence of CTAC (25 mM). Typically, a 100 mL 25 mM solution of CTAC was kept at 30 ºC, followed by addition of 200 µL of 100 mM ascorbic acid and 800 µL of a 1 mM dispersion of gold nanorods (corresponding to $[Au^{3+}]/[Au^0]$ = 12.5). After 15 min, 200 µL of a 50 mM $HAuCl_4$ solution was added to start the overgrowth process. After 90 min, the particles were centrifuged at 3,000 rpm for 10 min and redispersed in water. The dimensions of the final bipyramids were 182 nm (tip to tip) × 91 nm (at the equatorial plane).

### Sample and measurements Rotational tracking: 2D sample

Gold bipyramids (182 × 91 nm) were spin-coated onto ozonated coverslips and dried under vacuum for 3 h. Samples were sealed with a 120 µm imaging spacer and a second coverslip. A λ/2 waveplate was inserted between Lens 2 and the multimode cube, and the sample was mounted on a motorised precision rotation mount (PRM1Z8, Thorlabs) controlled by a KDC101 K-Cube DC Servo Motor via Kinesis software (Thorlabs).

### Sample and measurements: Poly(N-isopropylacrylamide) mixtures

Poly-N-isopropylacrylamide (~40,000 Mw; Sigma Aldrich) was dissolved at 10 mg/mL in Milli-Q water and mixed with FluoSpheres™ Carboxylate-Modified Microspheres (ThermoFisher, yellow-green, 1:500). 50 µL of the mixture was placed on VaHeat single slips (Interherence) or Grace Bio-Labs CoverWell™ perfusion chambers (0.6 mm thickness, 20 mm diameter). Temperature control was achieved with a VaHeat heating stage, ramping from 30 to 35°C at 0.5°C/min and cooling to 23°C, with measurements taken 3 min after stabilization.

### Autocorrelation analysis and rotational microrheology

Particle intensity was extracted using custom MATLAB code [10], with preprocessing including plane-wise intensity normalization. Local particle intensity was measured in a 5×5 pixel ROI, subtracting the 3-pixel outer background and summing the three brightest planes for each particle. In-plane rotation

($\vartheta$) was calculated from polarization ratio $r(t) = \frac{I_1 - I_2}{I_1 + I_2}$, and out-of-plane rotation from the total intensity $I_{Tot} = I_{Ch1} + I_{Ch2}$. Bot were detrended with a 100-point moving average, and autocorrelations $G(\tau)$ were obtained by MATLAB's xcorr function. The autocorrelation was fitted with single exponentials (best model via $\chi^2$ and BIC between single-, bi- and stretched exponential) to extract the lifetime $\tau_c$ from which diffusion coefficients were computed as $D_r = \frac{1}{n*\tau_c*2}$ with *n = 4* for 2D measurements (regarding one angle). The factor of 2 accounts for nanoparticle symmetry: an 180° rotation produces the same intensity pattern, meaning that each full particle rotation corresponds to two cycles in intensity. Out-of-plane rotation ($\phi$) was obtained from total intensity $I_{Tot} = I_1 + I_2$. Rotational microrheology-derived viscosities were estimated using the Broersma-type slender-body correction on Stokes-Einstein, and directed rotational motion was quantified from autocorrelation periodicity, $v = \frac{360°}{T}$, combining in- and out-of-plane components $v_r = \sqrt{v_\theta^2 + v_\varphi^2}$

**Acknowledgment**

This work was supported by the Flemish Government through long-term structural funding Methusalem (CASAS2, Meth/15/04), by the Fonds voor Wetenschappelijk Onderzoek Vlaanderen (FWO, W002221N), by a bilateral agreement between FWO and MOST (VS01925N), by the internal funds of KU Leuven (C1-project, C14/22/085), by the European Union under the Horizon Europe grant 101130615 (FASTCOMET), by the Spanish Agencia Estatal de Investigación and FEDER (PID2022-137569NA-C44)

B.L. acknowledges FWO for his junior postdoctoral fellowship (12AGZ24N). R.B.-O. thanks the Spanish Agencia Estatal de Investigación for a Ramon y Cajal contract (RYC2021-032773-I). S.H. thanks FWO for his junior postdoctoral fellowship (11A0S25N).

# References

1. S. T. Hess, T. P. K. Girirajan, and M. D. Mason, "Ultra-high resolution imaging by fluorescence photoactivation localization microscopy," Biophys. J. **91**, 4258–4272 (2006).
2. M. J. Rust, M. Bates, and X. Zhuang, "Sub-diffraction-limit imaging by stochastic optical reconstruction microscopy (STORM)," Nat. Methods **3**, 793–795 (2006).
3. S. W. Hell, "Far-field optical nanoscopy," 2010 IEEE Photinic Society's 23rd Annual Meeting (2010).
4. T. Dertinger, R. Colyer, G. Iyer, S. Weiss, and J. Enderlein, "Fast, background-free, 3D super-resolution optical fluctuation imaging (SOFI)," Proc. Natl. Acad. Sci. U. S. A. **106**, 22287–22292 (2009).
5. M. Dai, "DNA-PAINT Super-Resolution Imaging for Nucleic Acid Nanostructures," in *3D DNA Nanostructure*, Y. Ke and P. Wang, eds., Methods in Molecular Biology (Springer New York, 2017), Vol. 1500, pp. 185–202.
6. K. C. Gwosch, J. K. Pape, F. Balzarotti, P. Hoess, J. Ellenberg, J. Ries, and S. W. Hell, "MINFLUX nanoscopy delivers 3D multicolor nanometer resolution in cells," Nat Methods **17**, 217–224 (2020).
7. M. G. L. Gustafsson, "Surpassing the lateral resolution limit by a factor of two using structured illumination microscopy. SHORT COMMUNICATION," J. Microsc. **198**, 82–87 (2000).
8. M. Keshavarz, H. Engelkamp, J. Xu, E. Braeken, M. B. J. Otten, H. Uji-i, E. Schwartz, M. Koepf, A. Vananroye, J. Vermant, R. J. M. Nolte, F. De Schryver, J. C. Maan, J. Hofkens, P. C. M. Christianen, and A. E. Rowan, "Nanoscale Study of Polymer Dynamics," ACS Nano **10**, 1434–1441 (2016).
9. D. Wöll and C. Flors, "Super-resolution Fluorescence Imaging for Materials Science," Small Methods **1**, 1700191 (2017).
10. B. Louis, R. Camacho, R. Bresolí-Obach, S. Abakumov, J. Vandaele, T. Kudo, H. Masuhara, I. G. Scheblykin, J. Hofkens, and S. Rocha, "Fast-tracking of single emitters in large volumes with nanometer precision," Optics Express **28**, 28656–28656 (2020).
11. J. B. Pawley, "Fundamental Limits in Confocal Microscopy," in *Handbook Of Biological Confocal Microscopy*, J. B. Pawley, ed. (Springer US, 2006), pp. 20–42.
12. M. Badieirostami, M. D. Lew, M. A. Thompson, and W. E. Moerner, "Three-dimensional localization precision of the double-helix point spread function versus astigmatism and biplane," Appl. Phys. Lett. **97**, 161103 (2010).
13. S. R. P. Pavani, M. A. Thompson, J. S. Biteen, S. J. Lord, N. Liu, R. J. Twieg, R. Piestun, and W. E. Moerner, "Three-dimensional, single-molecule fluorescence imaging beyond the diffraction limit by using a double-helix point spread function," Proc. Natl. Acad. Sci. U. S. A. **106**, 2995–2999 (2009).
14. Y. Shechtman, L. E. Weiss, A. S. Backer, S. J. Sahl, and W. E. Moerner, "Precise Three-Dimensional Scan-Free Multiple-Particle Tracking over Large Axial Ranges with Tetrapod Point Spread Functions," Nano Lett. **15**, 4194–4199 (2015).
15. B.-C. Chen, W. R. Legant, K. Wang, L. Shao, D. E. Milkie, M. W. Davidson, C. Janetopoulos, X. S. Wu, J. A. Hammer 3rd, Z. Liu, B. P. English, Y. Mimori-Kiyosue, D. P. Romero, A. T. Ritter, J. Lippincott-Schwartz, L. Fritz-Laylin, R. D. Mullins, D. M. Mitchell, J. N. Bembenek, A.-C. Reymann, R. Böhme, S. W. Grill, J. T. Wang, G. Seydoux, U. S. Tulu, D. P. Kiehart, and E. Betzig, "Lattice light-sheet microscopy: imaging molecules to embryos at high spatiotemporal resolution," Science **346**, 1257998 (2014).
16. S. Abrahamsson, J. Chen, B. Hajj, S. Stallinga, A. Y. Katsov, J. Wisniewski, G. Mizuguchi, P. Soule, F. Mueller, C. D. Darzacq, X. Darzacq, C. Wu, C. I. Bargmann, D. A. Agard, M. Dahan, and M. G. L. Gustafsson, "Fast multicolor 3D imaging using aberration-corrected multifocus microscopy," Nat Methods **10**, 60–63 (2013).
17. S. Geissbuehler, A. Sharipov, A. Godinat, N. L. Bocchio, P. A. Sandoz, A. Huss, N. A. Jensen, S. Jakobs, J. Enderlein, F. Gisou van der Goot, E. A. Dubikovskaya, T. Lasser, and M. Leutenegger, "Live-cell multiplane three-dimensional super-resolution optical fluctuation imaging," Nat. Commun. **5**, 5830 (2014).

18. A. Descloux, K. S. Grußmayer, E. Bostan, T. Lukes, A. Bouwens, A. Sharipov, S. Geissbuehler, A.-L. Mahul-Mellier, H. A. Lashuel, M. Leutenegger, and T. Lasser, "Combined multi-plane phase retrieval and super-resolution optical fluctuation imaging for 4D cell microscopy," Nature Photon **12**, 165–172 (2018).
19. B. Louis, C.-H. Huang, R. Camacho, I. G. Scheblykin, T. Sugiyama, T. Kudo, M. Melendez, R. Delgado-Buscalioni, H. Masuhara, J. Hofkens, and R. Bresoli-Obach, "Unravelling 3D Dynamics and Hydrodynamics during Incorporation of Dielectric Particles to an Optical Trapping Site," ACS Nano **17**, 3797–3808 (2023).
20. F. de Jong, P. Diez-Silva, J.-K. Chen, R. Pérez-Peláez, S. Seth, H. Balakrishnan, B.-Y. Shih, M. Rosmeulen, S. Nonell, S. Rocha, A. Klymchenko, L. Liz-Marzán, R. Bresolí-Obach, M. I. Marqués, R. D. Buscalioni, J. Hofkens, and B. Louis, "Three-dimensional Optical Reconstruction of colloidal electrokinetics via multiplane imaging," (2025).
21. A. Gholivand, R. Bresolí-Obach, B. Louis, K. Dahlhoff, T. Dickscheid, J. Hofkens, and M. P. Lettinga, "Applying 8-foci imaging to instantaneous three-dimensional flow field reconstruction," Physics of Fluids **37**, 112010 (2025).
22. A. Dani, B. Huang, J. Bergan, C. Dulac, and X. Zhuang, "Superresolution Imaging of Chemical Synapses in the Brain," Neuron **68**, 843–856 (2010).
23. B. G. Kopek, G. Shtengel, J. B. Grimm, D. A. Clayton, and H. F. Hess, "Correlative Photoactivated Localization and Scanning Electron Microscopy," PLoS ONE **8**, e77209 (2013).
24. A. J. Hsieh, D. Veysset, D. F. Miranda, S. E. Kooi, J. Runt, and K. A. Nelson, "Molecular influence in the glass/polymer interface design: The role of segmental dynamics," Polymer **146**, 222–229 (2018).
25. S. Abrahamsson, M. McQuilken, S. B. Mehta, A. Verma, J. Larsch, R. Ilic, R. Heintzmann, C. I. Bargmann, A. S. Gladfelter, and R. Oldenbourg, "MultiFocus Polarization Microscope (MF-PolScope) for 3D polarization imaging of up to 25 focal planes simultaneously," Opt. Express **23**, 7734 (2015).
26. I. Gregor, E. Butkevich, J. Enderlein, and S. Mojiri, "Instant three-color multiplane fluorescence microscopy," Biophysical Reports **1**, 100001 (2021).
27. R. Thomas, I.-K. Park, and Y. Jeong, "Magnetic Iron Oxide Nanoparticles for Multimodal Imaging and Therapy of Cancer," IJMS **14**, 15910–15930 (2013).
28. E. M. Furst and T. M. Squires, *Microrheology* (Oxford University Press, 2017).
29. R. P. Mohanty and R. N. Zia, "The impact of hydrodynamics on viscosity evolution in colloidal dispersions: Transient, nonlinear microrheology," AIChE Journal **64**, 3198–3214 (2018).
30. A. J. Levine and T. C. Lubensky, "One- and Two-Particle Microrheology," Phys. Rev. Lett. **85**, 1774–1777 (2000).
31. G. Mattei, L. Cacopardo, and A. Ahluwalia, "Engineering Gels with Time-Evolving Viscoelasticity," Materials **13**, 438 (2020).
32. G. Hong, H. Tsuka, T. Maeda, Y. Akagawa, and K. Sasaki, "The dynamic viscoelasticity and water absorption characteristics of soft acrylic resin materials containing adipates and a maleate plasticizer," Dent. Mater. J. **31**, 139–149 (2012).
33. E. Schoolaert, P. Ryckx, J. Geltmeyer, S. Maji, P. H. M. Van Steenberge, D. R. D'hooge, R. Hoogenboom, and K. De Clerck, "Waterborne Electrospinning of Poly( *N* -isopropylacrylamide) by Control of Environmental Parameters," ACS Appl. Mater. Interfaces **9**, 24100–24110 (2017).
34. G. Scrofani, G. Saavedra, M. Martínez-Corral, and E. Sánchez-Ortiga, "Three-dimensional real-time darkfield imaging through Fourier lightfield microscopy," Opt. Express **28**, 30513 (2020).
35. M. J. Saxton, "Single-particle tracking: the distribution of diffusion coefficients," Biophysical Journal **72**, 1744–1753 (1997).
36. L. De Keer, K. I. Kilic, P. H. M. Van Steenberge, L. Daelemans, D. Kodura, H. Frisch, K. De Clerck, M.-F. Reyniers, C. Barner-Kowollik, R. H. Dauskardt, and D. R. D'hooge, "Computational prediction of the molecular configuration of three-dimensional network polymers," Nat. Mater. **20**, 1422–1430 (2021).
37. H. Sun, Z. Wang, and Y. He, "Direct Observation of Spatiotemporal Heterogeneous Gelation by Rotational Tracking of a Single Anisotropic Nanoprobe," ACS Nano **13**, 11334–11342 (2019).

38. J. F. Berret, "Microrheology of viscoelastic solutions studied by magnetic rotational spectroscopy," IJNT **13**, 597 (2016).
39. J. Michalski, T. Kalwarczyk, K. Kwapiszewska, J. Enderlein, A. Poniewierski, A. Karpińska, K. Kucharska, and R. Hołyst, "Rotational and translational diffusion of biomolecules in complex liquids and HeLa cells," Soft Matter **20**, 5810–5821 (2024).
40. N. C. Vu, Z. Ouzit, C. Lethiec, A. Maître, L. Coolen, F. Lerouge, and J. Laverdant, "Single Gold Bipyramid Nanoparticle Orientation Measured by Plasmon-Resonant Scattering Polarimetry," J. Phys. Chem. Lett. **12**, 752–757 (2021).
41. M. Philipp, R. Aleksandrova, U. Müller, M. Ostermeyer, R. Sanctuary, P. Müller-Buschbaum, and J. K. Krüger, "Molecular *versus* macroscopic perspective on the demixing transition of aqueous PNIPAM solutions by studying the dual character of the refractive index," Soft Matter **10**, 7297–7305 (2014).
42. P. J. Keller, A. D. Schmidt, J. Wittbrodt, and E. H. K. Stelzer, "Reconstruction of Zebrafish Early Embryonic Development by Scanned Light Sheet Microscopy," Science **322**, 1065–1069 (2008).
43. J. Li and Z. Xu, "Simultaneous dual-color light sheet fluorescence imaging flow cytometry for high-throughput marine phytoplankton analysis: retraction," Opt. Express **25**, 20033 (2017).
44. A. Chanda and C. Callaway, "Tissue Anisotropy Modeling Using Soft Composite Materials," Applied Bionics and Biomechanics **2018**, 1–9 (2018).
45. S. Ye, Y. Tong, A. Ge, L. Qiao, and P. B. Davies, "Interfacial Structure of Soft Matter Probed by SFG Spectroscopy," The Chemical Record **14**, 791–805 (2014).
46. A. Edelstein, N. Amodaj, K. Hoover, R. Vale, and N. Stuurman, "Computer Control of Microscopes Using μManager," CP Molecular Biology **92**, (2010).
47. A. D. Edelstein, M. A. Tsuchida, N. Amodaj, H. Pinkard, R. D. Vale, and N. Stuurman, "Advanced methods of microscope control using μManager software," JBM **1**, 1 (2014).